\documentclass[preprint,showpacs,preprintnumbers,amsmath,amssymb]{revtex4}%
\usepackage{amsmath}
\usepackage{mathrsfs}
\usepackage{graphicx}
\usepackage{dcolumn}
\usepackage{bm}

\begin{document}
\title{The Energy-Level Shifts of a Stationary Hydrogen Atom in Static External Gravitational Field
with Schwarzschild Geometry}

\author{Zhen-Hua Zhao$^{1,}$\footnote{zhaozhenhua@impcas.ac.cn},
Yu-Xiao Liu$^{2}$\footnote{liuyx@lzu.edu.cn}, Xi-Guo Lee$^{1}$ }

\affiliation{ $^{1}$Institute of Modern Physics, Chinese Academy
of Sciences, Lanzhou 730000, China\\  $^{2}$Institute of
Theoretical Physics, Lanzhou University, Lanzhou 730000, China}


\begin{abstract}
The first order perturbations of the energy levels of a stationary
hydrogen atom in static external gravitational field, with
Schwarzschild metric, are investigated. The energy shifts are
calculated for the relativistic $1S$, $2S$, $2P$, $3S$, $3P$,
$3D$, $4S$, $4P$, $4D$ and $4F$ levels. The results show that the
energy-level shifts of the states with total angular momentum quantum number
$1/2$ are all zero, and the ratio of absolute energy shifts with
total angular momentum quantum number $5/2$ is $1:4:5$. This feature can be used
to help us to distinguish the gravitational effect from other
effect.
\end{abstract}

\maketitle

\section{Introduction}

The study of gravitational fields interacting with spinor fields
constitutes an important element in constructing a theory that
combines quantum physics and gravity. For this reason, the
investigation of the behavior of relativistic particles in this
context is of considerable interest. With this interaction the
energy levels of an atom placed in external gravitational field will
be shifted. And these shifts, which would depend on the features of
the spacetime, are different for each energy level. So they can be
distinguished from the Doppler effect and from the gravitational and
cosmological red-shifts, in which cases the shifts would be the same
for all spectral lines. Thus the atomic spectra carries the
information about the local curvature at the position of the atom.
That may be led to a test of general relativity at the quantum
level, and one can, in principle, use the atom as an instrument to
detect possible regions of high curvature.

The hydrogen atom as the most simplest structure atom is an ideal
object to study this interaction and to detect the space-time
curvature\cite{Parker,Parker2,Parker3,Fischbach}. Using the Fermi
normal coordinates the energy-level shifts of hydrogen in a region
of curved space-time has been investigated in
Refs.~\cite{Parker,Parker2,Parker3}. The Fermi normal coordinates
require that the hydrogen atom is free falling along a
geodesic\cite{Parker,Parker2,Parker3}. The mixing of opposite-parity
states of an atom supported in gravitational field by
non-gravitational forces was investigated in Ref.~\cite{Fischbach}.
And that paper showed that the separation of center-of-mass and
relative coordinates is a complicated problem and the form of
Hamiltonian depends on the choice of center-of-mass and relative
coordinates. We think our calculation is not relevant to the
problem. Because, firstly, just like as the discussion in the last
paragraph of the page 2164 in Ref. \cite{Fischbach}, that the mass
of electron is much lighter than that of proton, so the problem can
be neglected with good approximate. Secondly, with the same approach
by Manasse and Misner in Ref. \cite{Manasse}, we think that the
separation of center-of-mass and relative coordinates has been
solved comfortably. Because with the method of Manasse and Misner,
the metric, connection, tetrad, etc., can be expanded with the
Riemann tensor at relative coordinates, and the Riemann tensor is
decided by the center-of-mass coordinate. So we think that we can
neglect that problem in our paper.

In this paper we also investigate the energy-level shifts of a
stationary hydrogen atom, with good approximately, in static
external gravitational field with Schwarzschild metric. In
particular, we use Riemann normal coordinates which are only normal
in a neighborhood of a space-time point. In this paper we prove the
equivalence of the gravitation effects of a hydrogen atom freely
falling along a radial geodesic \cite{Parker3} and resting in static
external gravitational field. We also redo the calculation in this
paper because in Ref. \cite{Parker3} the authors don't calculate the
parts of total angular momentum quantum number 5/2. And from our
calculation, we find that the ratio of absolute energy shifts of
those parts is 1:4:5, and the states with total angular momentum
quantum number $\frac{1}{2}$ are zero. We think these are good
results and can be used to help us to separate the shifts in the energy levels
caused by other effects.

In our earlier paper \cite{zhao} we have investigated the first
order perturbations of the energy levels of a hydrogen atom in
central internal gravitational field, which is produced by the mass
of the atomic nucleus. And the energy-level shifts are caused by
interaction of electronic with internal gravitational field. In this
paper we continue to study the energy-level shifts of a stationary
hydrogen atom, which are caused by the static external gravitational
field, with Schwarzschild metric. In this case we need to solve the
problem of the separation of center-of-mass and relative
coordinates, and which doesn't appear in our earlier paper
\cite{zhao}. And the energy-level shifts are not fixed in this case
with the uncertain of the mass of gravitation source and the
position of the atom.

In this paper we use the space-time signature ($ -,+,+,+$). And
under the transformations of coordinates Greek indices $\mu,\nu,
\ldots$, which imply values from 0 to 3, are regarded as the
tensor indices lowered with the curved space-time metric
$g_{\mu\nu}$. While indices $a,b,c,d$ (0 to 3) are the Lorentz
group indices `lowering' with flat space-time metric $\eta_{ab}$.

This paper is organized as follows. In Sec.~\ref{Dirac} we review
the formalism of the Dirac equation in curved space-time. In
Sec.~\ref{Riemann} metric and affinity are calculated in Riemann
normal coordinates. In Sec.~\ref{Riemann tensors} we give the
Riemann tensor in Riemann normal coordinates with Schwarzschild
geometry. The relativistic energy shifts of the atom are
calculated in Sec.~\ref{GRAVITATIONAL PERTURBATION}. Finally,
Sec.~\ref{conclusion} is the conclusion.

\section{Generally Covariant  Dirac Equation in Curved
Space-Time}\label{Dirac}

The generally covariant form of the Dirac equation
 in gravitational and electromagnetic fields
can be written as\cite{Fischbach,Duan}
\begin{eqnarray}
  \Gamma^\mu(\partial_\mu -\omega_\mu -i q A_\mu)\psi(x) +\frac{mc}{ \hbar}
  \psi(x)=0,\label{DiracEq2}
\end{eqnarray}
where $\Gamma^\mu$ is the generalized Dirac-Pauli matrices
\begin{eqnarray}
\Gamma_\mu(x)=b^{a}_{\;\;\mu}(x)\gamma_{a},\label{Gamma}
\end{eqnarray}
$\omega_\mu$ is the spinor connection defined as
\begin{eqnarray}
  \omega_\mu&=&\frac{1}{2} I_{\alpha\beta}(\nabla_\mu b^{a}_{\;\;\nu})b^{b}_{\;\;\lambda}
            g^{\lambda\nu},\label{omega}
\end{eqnarray}
$A_\mu$ is the electromagnetic vector potential in curved
space-time, $q=-e$ is the charge of electron, $b^{a}_{\;\;\mu}$ is
the tetrad (vierbein) field satisfying the relations
\begin{eqnarray}
  &&g_{\mu\nu}(x)=\eta_{ab}b^{a}_{\;\;\mu}(x)
  b^{b}_{\;\;\nu}(x),\label{tetrad1}\\
  &&b^{a}_{\;\;\mu}(x)b_{b}^{\;\;\mu}(x)=\delta^{a}_{b}. \label{tetrad2}
\end{eqnarray}
And $I_{ab}$ is the generator of $SO(3,1)$ group, whose spinor
representation is
\begin{eqnarray}
  I_{ab}= \frac{1}{4}(\gamma_a \gamma_b -\gamma_b \gamma_a).
\end{eqnarray}
Here $\gamma_a$ are the Dirac-Pauli matrices with the following
relation
\begin{eqnarray}
  \gamma_a \gamma_b +\gamma_b
  \gamma_a=2\eta_{ab},
\end{eqnarray}
and
\begin{eqnarray}
  \gamma_0^\dag&=&-\gamma_0,\quad \gamma_i^\dag=\gamma_i \qquad
  (i=1,2,3),\\
  \gamma_0&=&i \beta, \quad \quad  \gamma_i=-i
  \beta \alpha_i.\label{gamma}\\
  \alpha_{i}&=& \left(
  \begin{array}{cc}
   0&\;\sigma_i\\
   \sigma_i&\; 0
   \end{array} \right),\;
\beta= \left(
  \begin{array}{cc}
   I&\;0\\
   0&\; -I
\end{array} \right),
\end{eqnarray}
where $I$ is the  $2\times2$ identity  matrix,
and
$\sigma_i$ are the standard Pauli matrices
\begin{eqnarray}
  \sigma_1= \left(
  \begin{array}{cc}
   0&\; 1\\
   1&\; 0
\end{array} \right),\;
\sigma_2= \left(
  \begin{array}{cc}
   0&\; -i\\
   i&\; 0
\end{array} \right),\;
\sigma_3= \left(
  \begin{array}{cc}
   1&\; 0\\
   0&\; -1
\end{array} \right).
\end{eqnarray}

\section{Metric and Affinity in Riemann Normal Coordinates }\label{Riemann}

To calculate the energy-level shifts for a stationary hydrogen in
gravitational field, the Riemann normal coordinates is an ideal
mathematic tool. The well known Riemann normal coordinates satisfy
the conditions $g_{\mu\nu}|_{p_0}=\eta_{\mu\nu}|_{p_0}$ and
$\Gamma^\mu_{\nu\lambda}|_{p_0}=0$ at a point $p_0$---the origin
of the coordinate system. For arbitrary point $p$ in some
neighborhood $V(p_0)$  there is the unique geodic $\gamma(u)$
connecting $p_0$ and $p$, with $u$ being the canonical parameter.
The Riemann normal coordinates defined as $x^\mu=u\xi^\mu$, where
$\xi^\mu=dx^\mu/du$, with the origin at the point $p_0$. And a set
of orthonormal vectors $\vec{e}_0, \vec{e}_1, \vec{e}_2,
\vec{e}_3$ at the point $p_0$ fix the coordinate axes there. We
assume that the hydrogen atom is stationary with good
approximation at the space-time point $p_0$. So the Riemann normal
coordinates are also stationary  reference coordinates. Because
the gravitational field we considered is static so the derivative
with time is zero for all tensors and connections in Riemann
normal coordinates. The first order derivatives of affine
connection at the point $p_0$ are \cite{Manasse,Eisenhart}
\begin{subequations}
\begin{eqnarray}
&&\stackrel{0~}{\Gamma^\mu}_{\nu\lambda,0}=0,\\
&&\stackrel{0~}{\Gamma^\mu}_{0\nu,l}= \stackrel{0~}{\Gamma^\mu}_{\nu
0,l}=\stackrel{0~}{R^\mu}_{\nu 0 l},\\
&&\stackrel{0~}{\Gamma^\mu}_{ij,l}=
\frac{1}{3}(\stackrel{0~}{R^\mu}_{i j
l}+\stackrel{0~}{R^\mu}_{jil}),
\end{eqnarray}
\end{subequations}
where
$\stackrel{0~}{\Gamma^\mu}_{\nu\lambda,\rho}={\Gamma^\mu}_{\nu\lambda,\rho}(p_0)$.
Above formulas can be also got in Fermi normal
coordinates\cite{Manasse}, for the equation
$\Gamma^\mu_{\nu\lambda}|_G=0$ holds for all $x^0=\tau$ at the
timelike geodesic $G(\tau)$ in that frame. With the relation of
\begin{eqnarray}
\stackrel{0}{g}_{\mu\nu,ab}=
\eta_{\mu\sigma}\stackrel{0~}{\Gamma^\sigma}_{\nu a,b}+
\eta_{\sigma\nu}\stackrel{0~}{\Gamma^\sigma}_{\mu a,b},
\end{eqnarray}
the metric up to second order in these coordinates takes the form
\begin{subequations}\label{Riemann metric}
\begin{eqnarray}
&&g_{00}=-1+\stackrel{0}R_{0l0m}x^l x^m,\\
&&g_{0j}=g^{0j}=\frac{2}{3}\stackrel{0}R_{0ljm}x^l x^m\\
&&g_{ij}=\delta_{ij}+\frac{1}{3} \stackrel{0}R_{i ljm}x^l x^m,\\
&&g^{00}=-1-\stackrel{0~~~~~}{R^{0\;\;0}_{\;\;l\;\;m}} x^l x^m,\\
&&g^{ij}=\delta^{ij}-\frac{1}{3}
\stackrel{0~~~~~}{{R}^{i\;\;j}_{\;\;l\;\;m}} x^l x^m.
\end{eqnarray}
\end{subequations}
Using the equation of parallel propagation\cite{Nesterov} in the
Riemann normal coordinates
\begin{eqnarray}
\frac{d
b^a_{\;\;\mu}}{du}-\Gamma^{\lambda}_{\;\;\mu\sigma}b^a_{\;\;\lambda}\xi^\sigma=0,
\end{eqnarray}
we can write the tetrad as \cite{Parker2,Nesterov}
\begin{subequations}\label{Riemann tetrad}
\begin{eqnarray}
&&{b^a}_{0}=\delta^a_{0}+\frac{1}{2}\stackrel{0~}{\Gamma^a}_{0l,m}x^lx^m
=\delta^a_{0}+\frac{1}{2}\stackrel{0~}{R^a}_{\nu 0 l}x^lx^m,\\
&&{b_0}^{\mu}=\delta^\mu_{0}-\frac{1}{2}\stackrel{0~}{\Gamma^\mu}_{0l,m}x^lx^m
=\delta^\mu_{0}-\frac{1}{2}\stackrel{0~}{R^\mu}_{\nu 0 l}x^lx^m,\\
&&{b^a}_{i}=\delta^a_{i}+\frac{1}{2}\stackrel{0~}{\Gamma^a}_{il,m}x^lx^m
=\delta^a_{i}+\frac{1}{6}(\stackrel{0~}{R^a}_{i m
l}+\stackrel{0~}{R^a}_{mil})x^lx^m=\delta^a_{i}+\frac{1}{6}\stackrel{0~}{R^a}_{mil}x^lx^m,\\
&&{b_i}^{\mu}=\delta^\mu_{i}-\frac{1}{2}\stackrel{0~}{\Gamma^\mu}_{il,m}x^lx^m
=\delta^\mu_{i}-\frac{1}{6}(\stackrel{0~}{R^\mu}_{i m
l}+\stackrel{0~}{R^\mu}_{mil})x^lx^m=\delta^\mu_{i}-\frac{1}{6}\stackrel{0~}{R^\mu}_{mil}x^lx^m.
\end{eqnarray}
\end{subequations}
The affine connection components are
\begin{subequations} \label{Riemann connection}
\begin{eqnarray}
&&{\Gamma^\mu}_{0\nu}={\Gamma^\mu}_{\nu 0}=\stackrel{0~}{R^\mu}_{\nu 0 l} x^l,\\
&&{\Gamma^\mu}_{ij}=\frac{1}{3}(\stackrel{0~}{R^\mu}_{i j
l}+\stackrel{0~}{R^\mu}_{jil}) x^l,
\end{eqnarray}
\end{subequations}
where we are working to first order in $\stackrel{0~}{R^\mu}_{\nu
\lambda\rho}$. Using Eq. (\ref{omega}), the spinor connection
components are found to be
\begin{subequations}   \label{Riemann omega}
\begin{eqnarray}
\omega_0=-\frac{1}{2}\gamma_0\gamma_i\stackrel{0~}{R^{0i}}_{0l}x^l
         -\frac{1}{4}\gamma_i\gamma_j\stackrel{0~}{R^{ij}}_{0l}x^l,\\
\omega_k=-\frac{1}{4}\gamma_0\gamma_i\stackrel{0~}{R^{0i}}_{kl}x^l
         -\frac{1}{8}\gamma_i\gamma_j\stackrel{0~}{R^{ij}}_{kl}x^l,
\end{eqnarray}
\end{subequations}
working to first order in $\stackrel{0~}{R^\mu}_{\nu \lambda\rho}$.

We find that all above formulas have the same form as those in the Fermi
normal coordinates \cite{Parker2}, if the gravitational field is
chosen static. This can be regarded as a kind of equivalence
between Riemann normal coordinates and Fermi normal coordinates.

\section{Riemann Tensor in Riemann Normal Coordinates with Schwarzschild Geometry}\label{Riemann tensors}

In Sec. \ref{Riemann} we find in Riemann normal coordinates that
all physical quantities are expressed by the Riemann tensor at the
origin point $p_0$. So it is necessary to get the expression of
the Riemann tensor $\stackrel{0~}{R^\mu}_{\nu \lambda\rho}$ in
this coordinates. Therefore, we need another coordinates to
describe the external gravitational field. In Schwarzschild
coordinates, which we will call $X^{\mu'}$ or $cT,R,\Theta,\Phi,$
the metric components $g_{\mu'\nu'}$ are displayed in the term
\begin{eqnarray}\label{metric}
ds^2&=&-g_{\mu'\nu'}dX^{\mu'}dX^{\nu'}\nonumber\\
&=&c^{2} X dT^2 -\frac{1}{X}dR^2 -R^{2}d\Theta^2 -R^2 \sin^{2}\Theta
d\Phi^2,
 \end{eqnarray}
where $X=1-R_{S}/R, R_{S}=2GM/c^2$ is the Schwarzschild radius, $M$
is the mass of the gravitation source.

The next step in constructing Riemann normal coordinates is to
choose an orthonormal  frame at the point $p_0$. The timelike base
vector must be the tangent $\partial/\partial t$, and the symmetry
determines the others. Conveniently we choose
\begin{equation}\label{base vectors}
\begin{split}
&\vec{e}_0=\partial/\partial t=1/\sqrt{X}\;\partial/\partial T,\\
&\vec{e}_1=\partial/\partial
x=\sqrt{X}\;\partial/\partial R,\\
&\vec{e}_2=\partial/\partial y=1/R\; \partial/\partial
\Theta,\\
&\vec{e}_3=\partial/\partial z=1/(R\sin\Phi)\;\partial/\partial\Phi.
\end{split}
\end{equation}
Where $x,y,z,t$ are the Riemann normal coordinates. We can compute
the curvatures in the Riemann frame by the tensor transformation law
\begin{eqnarray}
R_{abcd}=R_{\mu'\nu'\lambda'\rho'}{e_\alpha}^{\mu'}{e_b}^{\nu'}{e_c}^{\lambda'}{e_d}^{\rho'}.
\end{eqnarray}
The curvature components $R_{\mu'\nu'\lambda'\rho'}$ with respect to
the Schwarzschild frame are well known as
\begin{equation}
\begin{split}
&R_{1'0'1'0'}=R_{S}/R^3,\;R_{2'0'2'0'}=-R_{S} X/(2R),\\
&R_{3'0'3'0'}=-R_{S} X\sin^2\Theta/(2R),\;R_{1'2'1'2'}=R_{S}/(2RX),\\
&R_{2'3'2'3'}=R_{S}R\sin^2\Theta,\;R_{1'3'1'3'}=R_{S}\sin^2\Theta/(2XR).
\end{split}
\end{equation}
Only the independent non-vanishing components are listed above and
below. From Eq.(\ref{base vectors}) we have
\begin{eqnarray}
[{e_a}^{\mu}]=\left(
\begin{array}{cccc}
 1/\sqrt{X} & 0 & 0 & 0 \\
 0 & \sqrt{X} & 0 & 0 \\
 0 & 0 & 1/R & 0 \\
 0 & 0 & 0 & 1/(R\sin\Theta)
\end{array}
\right).
\end{eqnarray}
The computation then yields
\begin{equation}
\begin{split}
&R_{1010}=\frac{R_{S}}{R^3},\;R_{1212}=R_{1313}=\frac{R_{S}}{2 R^3},\\
&R_{2020}=R_{3030}=-\frac{R_{S}}{2
R^3},\;R_{2323}=-\frac{R_{S}}{R^3}.
\end{split}
\end{equation}
The above expressions of Riemann tensor have the same form with
those in Fermi normal coordinates, which move along a radial
geodesic\cite{Parker3,Manasse}.


\section{Energy-level Shifts in the Schwarzschild Spacetime} \label{GRAVITATIONAL PERTURBATION}

Here we calculate the relativistic energy shifts of a hydrogen atom
in the Schwarzschild spacetime. The exact solutions of the Dirac
equation for a hydrogen atom in flat space-time serve as the basis
for perturbation theory.

The energy eigenvalues of a hydrogen atom are
\begin{eqnarray}
  E_{n\kappa}=m
  c^2/\sqrt{1+\left(\frac{\zeta}{n-|\kappa|+s}\right)^2},
\end{eqnarray}
where $\zeta=Ze^2$, $s=\sqrt{\kappa^2-\zeta^2}$, $m$ is the mass of
electron, $n=1,2,\cdots$ is the principal quantum number.

The bound state functions of a hydrogen atom can be written in
standard representation as\cite{Rose,Strange,zhao}
\begin{eqnarray}
  \psi=\psi^{M}_{\kappa}=\left( \begin{array}{c}g(r)\chi^{M}_{\kappa}\\
  -i f(r)\chi^{M}_{-\kappa}\end{array}\right),
\end{eqnarray}
here $M$ and $\kappa$ are the eigenvalues of $J_{z}$ and
$K=b(\vec{\sigma}\cdot\vec{L}+I)$, respectively. The functions
$f(r)$, $g(r)$ and spinors $\chi^M_\kappa$, $\chi^M_{-\kappa} $ are
given by
\begin{equation}
\begin{split}
f(r)&=\frac{2^{s-\frac{1}{2}}\lambda^{s+\frac{3}{2}}}{\Gamma(2s+1)}\sqrt{\frac{\Gamma(2s+n_r+1)}{n_r!
\zeta K_c(\zeta K_c-\lambda\kappa )}}
\sqrt{1-\frac{W_{c}}{K_c}}r^{s-1}e^{-\lambda r} \\
&\left(\left(\kappa -\frac{\zeta K_c}{\lambda} \right)\text{F}(-n_r,
2 s+1,2\lambda r)-n_r \text{F}(-n_r+1,2 s+1, 2\lambda r)\right),
\end{split}
\end{equation}
\begin{equation}
\begin{split}
g(r)&=-\frac{2^{s-\frac{1}{2}}\lambda^{s+\frac{3}{2}}}{\Gamma(2s+1)}\sqrt{\frac{\Gamma(2s+n_r+1)}{n_r!
\zeta K_c(\zeta K_c-\lambda\kappa )}}
\sqrt{1-\frac{W_{c}}{K_c}}r^{s-1} e^{-\lambda r} \\
&\left(\left(\kappa -\frac{\zeta K_c}{\lambda} \right)\text{F}(-n_r,
2 s+1,2\lambda r)+n_r \text{F}(-n_r+1,2 s+1, 2\lambda r)\right),
\end{split}
\end{equation}
\begin{equation}
\chi^M_\kappa=C_{1/2}Y^{M-1/2}_{l}\left(
  \begin{array}{c}
   1\\
   0
\end{array} \right)+C_{-1/2}Y^{M+1/2}_{l}\left(
  \begin{array}{c}
   0\\
   1
\end{array} \right),
\end{equation}
\begin{eqnarray}
 \chi^M_{-\kappa}=-C_{1/2}Y^{M-1/2}_{l}\left(
  \begin{array}{c}
   \cos\theta\\
   \sin\theta e^{i\phi}
\end{array} \right)-C_{-1/2}Y^{M+1/2}_{l}\left(
  \begin{array}{c}
   \sin\theta e^{-i\phi}\\
   -\cos\theta
\end{array} \right),
\end{eqnarray}
where $W_c=E_{n\kappa}/\hbar c$, $K_c=mc^2/\hbar c$,
$\lambda=\sqrt{m^2c^4-E_{n\kappa}^2}/\hbar c$, $\Gamma(2s+1)$ is the
$\Gamma$ function, $C_{1/2}$ and $C_{-1/2}$ are the C-G
coefficients.

The gravitational perturbation matrix elements are
\begin{eqnarray}
\langle H_{I}\rangle_{ij}\equiv(\psi_{i},H_{I}\psi_{j}).
\end{eqnarray}
Also with the analyzing in Ref.~\cite{Parker2}, the perturbation
Hamiltonian retaining the main term can be written as
\begin{eqnarray}
H_I=\frac{1}{2}\beta
     mc^2\stackrel{0~}{R^0}_{l0m}x^lx^m.
\end{eqnarray}
Using the equation\cite{Parker2}
\begin{eqnarray}
\text{det}[(\psi_i,H_{I}\psi_j)-E^{\;1}\delta_{ij}]=0\label{Ei},
\end{eqnarray}
and from the usual perturbation theory of a degenerate energy
eigenvalue, the relativistic energy shifts of the atom we calculated
can be found in Table~\ref{table1}.

\begin{table}[h]
\begin{ruledtabular}

\caption{\label{table1}The energy-level shifts }
{\begin{tabular}{@{}cc@{}}
State &    The energy-level shifts \\
\colrule
$1S_{1/2}$&0\\
\colrule
$2S_{1/2}$&0 \\
\colrule
$2P_{1/2}$& 0 \\
\colrule
$2P_{3/2}$& $\pm 5.80782\times10^{-72}Mm^{-1}R^{-3}$\\
\colrule
$3S_{1/2}$& 0\\
\colrule
$3P_{1/2}$& 0\\
\colrule
$3P_{3/2}$& $ \pm 1.74235\times10^{-71}M m^{-1}R^{-3}$\\
\colrule
$3D_{3/2}$& $\pm 6.96938\times10^{-72}M m^{-1}R^{-3}$\\
\colrule
$3D_{5/2}$&$-6.96944\times10^{-72}M m^{-1}R^{-3}$\\
          &$-2.72778\times10^{-71}M m^{-1}R^{-3}$\\
          &$3.48472\times10^{-71}M m^{-1}R^{-3}$\\
\colrule
$4S_{1/2}$& 0\\
\colrule
$4P_{1/2}$& 0\\
\colrule
$4P_{3/2}$& $\pm 4.18166\times10^{-71}M m^{-1}R^{-3}$\\
\colrule
$4D_{3/2}$& $\pm 2.32314\times10^{-71}M m^{-1}R^{-3}$\\
\colrule
$4D_{5/2}$&$-1.54877\times10^{-71}M m^{-1}R^{-3}$\\
          &$-6.19507\times10^{-71}M m^{-1}R^{-3}$\\
          &$7.74384\times10^{-71}M m^{-1}R^{-3}$\\
\colrule
$4F_{5/2}$&$-7.52257\times10^{-72}M m^{-1}R^{-3}$\\
          &$-3.00903\times10^{-71}M m^{-1}R^{-3}$\\
          &$3.72129\times10^{-71}M m^{-1}R^{-3}$\\
\end{tabular}}
\end{ruledtabular}
\end{table}

\section{Conclusion}\label{conclusion}
In this paper we investigate energy-level shifts of a stationary
hydrogen atom in static external gravitational field. With the
calculations we find the results in this case is same with a
hydrogen atom freely falling along a radical geodesic\cite{Parker3}.
From the results in Table~\ref{table1} we find that the energy
shifts of the states with total angular momentum quantum number
$\frac{1}{2}$ are zero, and the ratio of absolute energy shifts with
total angular momentum quantum number $\frac{5}{2}$ is $1:4:5$. This
feature can be used to help us to separate the shifts in the energy
levels caused by other effects.

\section{Acknowledgments}
This work was supported by the National Natural Science Foundation
of China under Grant Nos. 10435080 and 10575123, Chinese Academy of
Sciences Knowledge Innovation Project under Grant Nos. KJCX3.SYW.N2
and KJCX2-SW-N16, and the Foundational Reserach Fund for Physics and
Mathematic of Lanzhou University under Grant No. Lzu07002.

\end{document}